\documentclass[superscriptaddress,aps,prl,reprint,twocolumn,amsmath,amssymb]{revtex4-2}
\usepackage{graphicx}
\usepackage{slashed}
\usepackage{dcolumn}
\usepackage{bm}
\usepackage{multirow}
\usepackage{dsfont}
\usepackage{xcolor}

\usepackage{hyperref}
\hypersetup{
    colorlinks=true,
    linkcolor=blue,
    citecolor=blue,     
    urlcolor=blue,
}

\allowdisplaybreaks

\makeatletter

\newcommand{\Rmnum}[1]{\expandafter\@slowromancap\romannumeral #1@}
\makeatother
\begin{document}
\preprint{APS/123-QED}

\title{\textbf{Anomalous Piezoelectricity from Polarization-Dependent Electrostriction in Wurtzites} 
}

\author{Guo-Dong Zhao}
 \affiliation{
 Department of Materials Science and Engineering, Pennsylvania State University, University Park, Pennsylvania 16802, USA
}

\author{Aiden Ross}
 \affiliation{
 Department of Materials Science and Engineering, Pennsylvania State University, University Park, Pennsylvania 16802, USA
}

\author{Yueze Tan}
 \affiliation{
 Department of Materials Science and Engineering, Pennsylvania State University, University Park, Pennsylvania 16802, USA
}

\author{Yujie Zhu}
 \affiliation{
 Department of Materials Science and Engineering, University of Wisconsin-Madison, Madison, Wisconsin 53706, USA
}

\author{Jia-Mian Hu}
 \affiliation{
 Department of Materials Science and Engineering, University of Wisconsin-Madison, Madison, Wisconsin 53706, USA
}

\author{Long-Qing Chen}
 \email{lqc3@psu.edu}
 \affiliation{
 Department of Materials Science and Engineering, Pennsylvania State University, University Park, Pennsylvania 16802, USA
}

\date{\today}

\begin{abstract}
The piezoelectric coefficient is a third-rank tensor connecting the strain or stress with the electric field or polarization, whereas the electrostriction coefficient is a fourth-rank tensor relating the strain to the square of electric polarization.
The electrostriction tensor components in the current literature are often treated as constants independent of polarization, resulting in piezoelectric tensor components that are linearly proportional to polarization and the dielectric susceptibility tensor. 
Here, we study the electrostriction and piezoelectricity in strongly polar wurtzites, including AlN, Al$_{1-x}$Sc$_x$N, Al$_{1-x}$B$_x$N, GaN, and ZnO. 
We discover that electrostriction and the elastic modulus in wurtzites are both strongly polarization-dependent, and the piezoelectric coefficient is highly nonlinear with respect to polarization, including the anomalous possibility that decreasing polarization increases the electromechanical strain response. 
These unusual dependencies of electrostriction and piezoelectric effects on polarization arise from the evolution of a layered reference nonpolar structure toward a tetrahedrally coordinated wurtzite network structure as the polarization increases. 
The findings have important implications in understanding the thermodynamics of the general class of wurtzite ferroelectrics and in manipulating their piezoelectric and ferroelectric behaviors.
\end{abstract}

\maketitle

\textit{Introduction.---}
Piezoelectric materials underpin a wide range of modern technologies, including sensors, actuators, radio-frequency filters, and acoustic devices. 
Wurtzite-structured AlN and related materials are key platforms for bulk acoustic wave resonators and integrated electromechanical devices, owing to their high quality factors and excellent thermal and chemical stability~\cite{Li2025}. 
Alloying and strain engineering have further expanded this material class,
enabling ferroelectricity and strongly composition-tunable piezoelectricity in systems such as Al$_{1-x}$Sc$_x$N, Al$_{1-x}$B$_x$N, and Zn$_{1-x}$Mg$_x$O~\cite{Fichtner2019,Kim2023,Casamento2024,Zhang2024,Skidmore2025,Wang2025,Lee2024,Mercado2026}.
These developments open new opportunities for high-performance piezoelectric devices, and raise a basic question: what thermodynamic and microscopic factors control electromechanical coupling in polar wurtzites?

The piezoelectric strain response is often written as $d=2QP\varepsilon_0\chi$~\cite{Sundar1992,Newnham1997,Li2014}, where $Q$ is the electrostrictive coefficient, $P$ the polarization, $\varepsilon_0$ the vacuum permittivity, and $\chi$ the dielectric susceptibility.
This expression assumes that the stress-free strain is quadratic in polarization, $\varepsilon^0\simeq QP^2$, so that the fixed-stress strain-polarization factor is $b=(\partial\varepsilon^0/\partial P)_\sigma\simeq2QP$.
Enhancement of $d$ is then often discussed in terms of increasing $\chi$ through dielectric softening, i.e., flattening the polarization free-energy curvature, while keeping $P$ and $Q$ sizable~\cite{Park2022,Li2022,Yan2022}.
This description is useful when the strain-polarization relation remains close to quadratic, as in many conventional perovskite ferroelectrics, yet questions remain about this description in strongly polar ferroelectrics.

Polar wurtzites break this description because their large polar distortions drive substantial structural changes as polarization develops. 
As a result, the piezoelectric response is governed by a strongly polarization-dependent local strain-polarization coupling rather than by the conventional constant-$Q$ relation $b=2QP$. 
This structural renormalization may generate an additional contribution to $b$, establishing a self-consistent feedback between polarization and strain that is analogous to the self-consistent exchange field generated by spin polarization in Hartree-Fock theory~\cite{auerbach2012interacting,PhysRevB.92.155414,PhysRevB.93.235433}. 
Here, using AlN as the central example, we construct density functional theory (DFT) free-energy surfaces to determine the polarization dependences of the stress-free strain, electrostrictive coefficient, piezoelectric response, and elastic modulus.
The results demonstrate that the constant-electrostriction approximation breaks down for wurtzite AlN, leading to an overestimation of the longitudinal piezoelectric response by more than a factor of three, whereas the perovskite PbTiO$_3$ remains much closer to the conventional quadratic strain-polarization picture.

\textit{First-principles free-energy surface.---}
For a proper ferroelectric, we use the order-parameter polarization $P_i^{\rm s}$ and denote the stress-free strain by $\varepsilon^0_{ij}(P_k^{\rm s})$.
The total strain is decomposed as
$\varepsilon_{ij} \!=\! \varepsilon^{\rm el}_{ij} \!+\! \varepsilon^0_{ij}(P_k^{\rm s})$,
where $\varepsilon^{\rm el}_{ij}$ is the elastic strain relative to the stress-free branch.
At $T\!=\!0$~K, the chemo-mechanical free-energy density under a macroscopic electric field is~\cite{Chen2008}
\begin{align}
    f_{\rm CM}
    &=\! g_0 
     \!+\! \Delta g(P_i^{\rm s}) 
     \!+\! U_{\rm elas}\!\left(P_k^{\rm s},\varepsilon^{\rm el}_{ij}\right) 
     \!-\! E_i P_i^{\rm s}
     \!+\! \phi(E_i),
     \label{eqn:fCM}
\end{align}
where $g_0$ is the zero-field energy density of the unstrained nonpolar reference state, $\Delta g$ the polar-order contribution, $U_{\rm elas}$ the elastic energy measured relative to $\varepsilon^0_{ij}(P_k^{\rm s})$, and 
$\phi(E_i)$ collects field-only terms independent of $P_i^{\rm s}$ and $\varepsilon^{\rm el}_{ij}$~\cite{Tagantsev1986}.
The zero-field energy surface obtained from DFT is identified with the field-independent part of $f_{\rm CM}$,
\begin{align}
    f_{\rm DFT}(P_i^{\rm s},\varepsilon_{ij})
    =
    g_0 
    +
    \Delta g(P_i^{\rm s})
    +
    U_{\rm elas}\!\left(P_i^{\rm s},\varepsilon^{\rm el}_{ij}\right).
    \label{eqn:fDFTmap}
\end{align}
The equilibrium condition with respect to $f_{\rm CM}$ with respect to $P_i^{\rm s}$ at fixed total strain and applied field gives
$\left(\partial f_{\rm DFT}/\partial P_i^{\rm s}\right)_{\varepsilon}=E_i$.
To lowest order in the elastic strain about $\varepsilon^0_{ij}(P_i^{\rm s})$,
$\sigma_{ij}=C^P_{ijkl}(P_i^{\rm s})\varepsilon^{\rm el}_{kl}$,
where $C^P_{ijkl}$ is the stiffness at fixed order parameter, corresponding to the
harmonic elastic energy
$U_{\rm elas} =\frac12 C^P_{ijkl}(P_i^{\rm s}) \varepsilon^{\rm el}_{ij}\varepsilon^{\rm el}_{kl}$~\cite{SM}.

To evaluate longitudinal electromechanical coefficients in AlN, we reduce $f_{\rm DFT}$ to an effective free-energy surface $f_{\rm eff}(P_3,\varepsilon_{33})$ along the symmetry-allowed polar distortion path parameterized by $P_3$.
Here $P_1 \!=\! P_2 \!=\! 0$, transverse strains are relaxed under $\sigma_\perp \!=\! \sigma_{11} \!=\!\sigma_{22} \!=\! 0$, and shear strains vanish by symmetry along this single-mode path.
For AlN, this path evolves from the centrosymmetric layered-hexagonal $P6_3/mmc$ reference to the polar $P6_3mc$ wurtzite structure~\cite{Dreyer2016,Tsipas2013,Ye2025,Cai2007,Mercado2026}.
At each $P_3$, imposing $\sigma_{33}=0$ determines $\varepsilon^0_{33}(P_3)$ and the stress-free energy.
The value of $P_3$ at the relaxed polar ground state on this branch is denoted by $P_s$.
Along this stress-free branch, the local longitudinal electrostrictive coefficient is defined as 
$Q_{3333}(P_3) \equiv (\mathrm{d}^2\varepsilon^0_{33}/\mathrm{d} P_3^2)/2$,
as a local curvature rather than a global quadratic fit.
Local derivatives of $f_{\rm eff}$ yield $d_{333}$ and $\tilde C^P_{3333} (\partial\sigma_{33}/\partial\varepsilon_{33})_{P_3,\sigma_\perp=0}$, where the tilde denotes relaxation of the transverse strains differing from the fixed-transverse-strain one.
Computational details are given in SM~\cite{SM,Kresse1996-1,Kresse1996-2,Perdew2008,King1993,Mercado2026}.

\begin{figure}[htbp]
\centering
\includegraphics[width=\columnwidth]{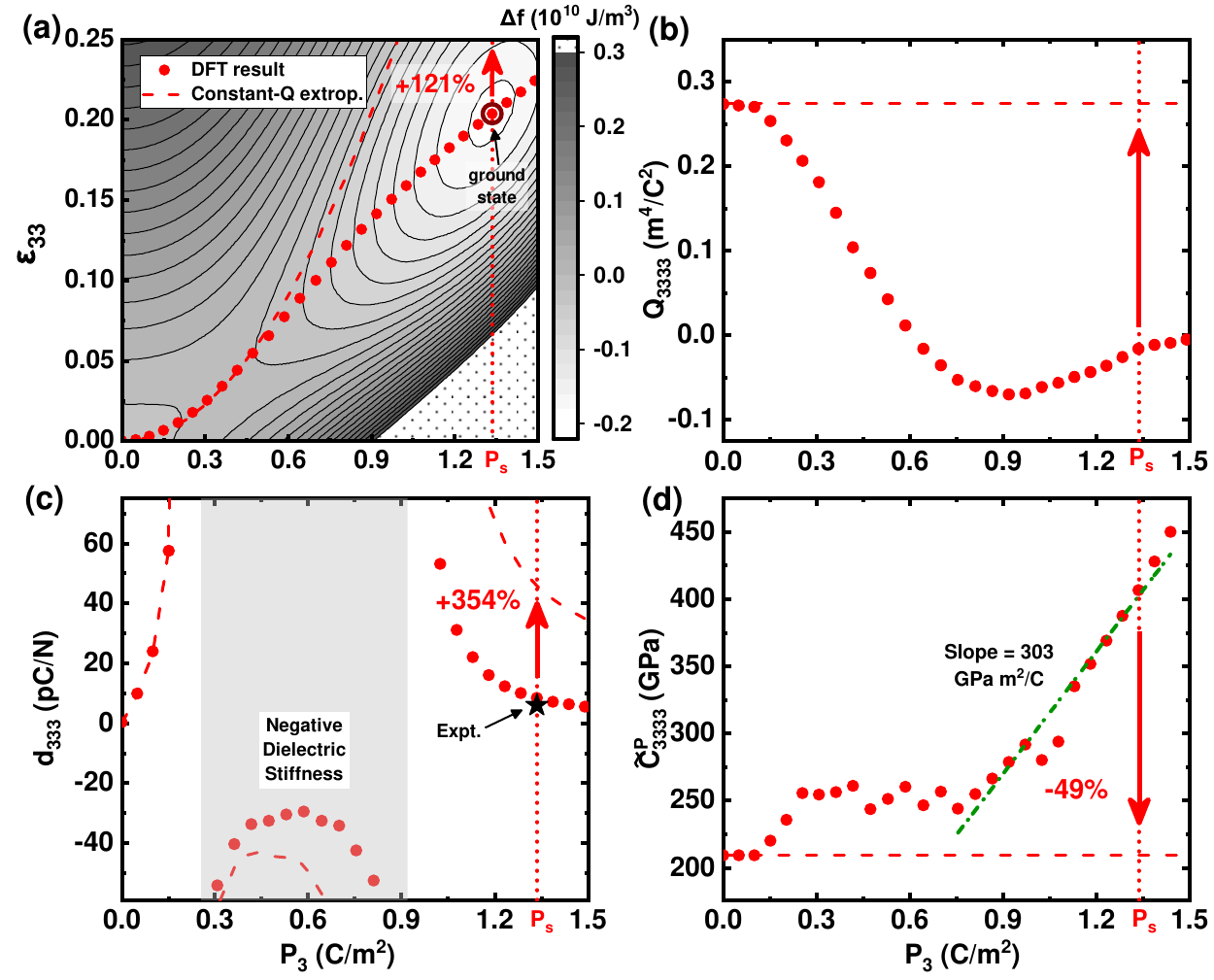}
\caption{Polarization-dependent electromechanical response in AlN.
    (a) $\Delta f(P_3,\varepsilon_{33})$ with the stress-free branch (red dots) and small-$P_3$ constant-$Q$ extrapolation (red dashed), overestimating $\varepsilon^0_{33}(P_s)$ by 121\%.
    (b) $Q_{3333}$ and its $P_3=0$ reference.
    (c) $d_{333}$ and its constant-$Q$ estimate, overestimating $d_{333}(P_s)$ by 354\%. 
    The black star marks experimental single-crystal value~\cite{Akiyama2009,Kim2015}.
    Shading marks the negative-stiffness region with $(\partial E_3/\partial P_3)_\sigma \!<\! 0$.
    (d) $\tilde C^P_{3333}$ and its $P_3 \!=\! 0$ reference, underestimating $\tilde C^P_{3333}(P_s)$ by 49\%. 
    The green dash-dotted line is a high-$P_3$ linear fit.
    Vertical dotted lines mark $P_s$, and red arrows indicate the quoted deviations from DFT values at $P_s$.
}
\label{fig:fig1}
\end{figure}

Fig.~\ref{fig:fig1}(a) shows the stress-free branch $\varepsilon^0_{33}(P_3)$ of AlN on the two-variable free-energy surface.
$\varepsilon^0_{33}$ is nearly quadratic at small $P_3$, but the constant-$Q$ extrapolation increasingly departs from the DFT strain branch at large polarization, overestimating $\varepsilon^0_{33}(P_s)$ by 121\%.
Thus, the quadratic strain-polarization relation obtained near $P_3\!=\!0$ does not extend to $P_s$.

\begin{figure}[htbp]
\centering
\includegraphics[width=\columnwidth]{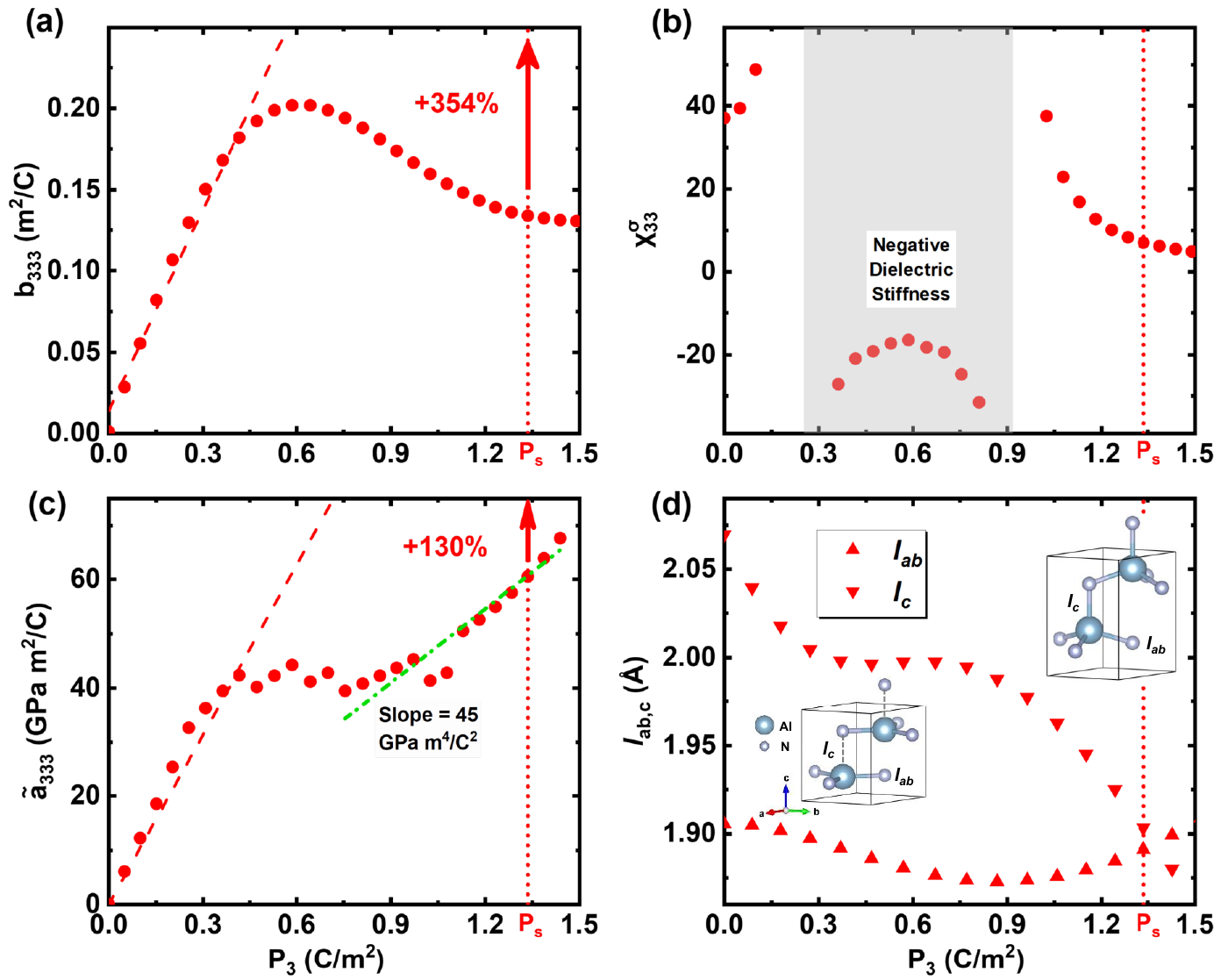}
\caption{Decomposition and structural origin of polarization-dependent piezoelectricity in AlN.
    (a) Fixed-stress strain-polarization factor $b_{333}$,
    (b) fixed-stress susceptibility $\chi^\sigma_{33}$. 
        Shading in (b) marks the negative-stiffness region,
    (c) electromechanical stress coupling $\tilde a_{333}$, and
    (d) in-plane and axial Al-N bond lengths $l_{ab}$ and $l_c$ in the layered and wurtzite structures.
    Red dashed curves are small-$P_3$ constant-$Q$ extrapolations, the green dash-dotted line in (c) is a near-$P_s$ fit, and vertical dotted lines mark $P_s$.
}
\label{fig:fig2}
\end{figure}

The local electrostrictive coefficient in Fig.~\ref{fig:fig1}(b) makes this failure explicit:
$Q_{3333}(P_3)$ decreases rapidly and becomes slightly negative near $P_s$.
Consequently, the constant-$Q$ estimate is reliable only near $P_3=0$ and overestimates $d_{333}(P_s)$ by 354\% [Fig.~\ref{fig:fig1}(c)].
The negative $d_{333}$ values in the shaded region originate from the sign change of the fixed-stress susceptibility $\chi^\sigma_{33}$ [Fig.~\ref{fig:fig2}(b)].
This shaded segment is an unstable negative-stiffness region, $(\partial E_3/\partial P_3)_\sigma \!<\! 0$, analogous to negative-permittivity/negative-capacitance regions~\cite{Khan2014,Zubko2016,Yadav2019,Jorge2019}.
At stress-free equilibrium, $d_{333}(P_s)=8$~pC/N at $0$~K, compares well with the room-temperature single-crystal value of about $6$~pC/N~\cite{Akiyama2009,Kim2015}.

Fig.~\ref{fig:fig1}(d) further shows that $\tilde C^P_{3333}$ increases strongly with increasing $P_3$ and is nearly linear above about $0.7~\mathrm{C/m^2}$.
Approximating $\tilde C^P_{3333}$ by its $P_3=0$ value over the full polarization range underestimates $\tilde C^P_{3333}(P_s)$ by 49\%.
Together, these trends motivate decomposing $d_{333}$ into the strain-polarization factor $b_{333}$ and the dielectric response $\chi^\sigma_{33}$ below.

\textit{Decomposition and structural origin of $d_{333}$.---}
For the local fixed-stress response evaluated along the stress-free branch, the chain rule gives the order-parameter-mediated converse piezoelectric response,
\begin{align}
    d_{ijk}
    \equiv \left( \frac{\partial\varepsilon_{ij}} {\partial E_k} \right)_\sigma
    &=
    \left( \frac{\partial \varepsilon^0_{ij}}{\partial P_m^s} \right)_{\sigma}
    \left( \frac{\partial P^s_m}{\partial E_k} \right)_{\sigma}
    =
    b_{ijm}\varepsilon_0\chi^\sigma_{mk},
    \label{eqn:d}
\end{align}
where $b_{ijm}\equiv (\partial\varepsilon^0_{ij}/\partial P_m^s)_\sigma$ and $\varepsilon_0\chi^\sigma_{mk} \equiv (\partial P_m^s/\partial E_k)_\sigma$~\cite{Cady1946,Haskins1950,Bechmann1953}.
Thus, for the longitudinal response, $d_{333}=b_{333}\varepsilon_0\chi^\sigma_{33}$.
After eliminating the transverse strains under zero transverse stress, the reduced longitudinal stress variation is
\begin{align}
    \mathrm{d}\sigma_{33}
    &=
    \tilde C^P_{3333}\,\mathrm{d}\varepsilon_{33}
    -\tilde a_{333}\,\mathrm{d}P_3.
\end{align}
Imposing $\mathrm{d}\sigma_{33}=0$ gives
\begin{align}
    b_{333}
    \equiv 
    \left( \frac{\partial \varepsilon^0_{33} }{\partial P_3 }\right)_{\sigma}
    &=
    S^P_{3333}\tilde a_{333}
    \label{eqn:dsplit}
\end{align}
where $S^P_{3333}\equiv(\tilde C^P_{3333})^{-1}$ is the reduced longitudinal compliance.
Here $\tilde a_{333}\equiv -\partial^2f_{\rm eff}/ (\partial\varepsilon_{33}\partial P_3)$ 
is the mixed strain-polarization coupling of the reduced surface. 
The unreduced matrix of second derivatives and the transverse-stress reduction are given in the SM~\cite{SM}.

Fig.~\ref{fig:fig2}(a) shows that $b_{333}$ itself turns over near $P_s$, whereas the constant-$Q$ expectation would keep increasing linearly with $P_3$.
Counterintuitively, reducing $P_3$ from $P_s$ can therefore increase the electromechanical factor $b_{333}$, suggesting a route to enhance the strain-polarization contribution to $d_{333}$.
Although $\chi^\sigma_{33}$ represents the polarization-response factor in $d_{333}$~\cite{Hao2026,Budimir2004,Budimir2006}, Fig.~\ref{fig:fig2}(a) shows that $b_{333}$ cannot be replaced by $2Q^{(0)}P_3$, where $Q^{(0)}\equiv Q_{3333}(0)$ is the small-$P_3$ curvature.

The reduced stress coupling $\tilde a_{333}\equiv -\partial^2 f_{\rm eff}/(\partial\varepsilon_{33}\partial P_3)$ also varies nonlinearly with $P_3$ [Fig.~\ref{fig:fig2}(c)]:
a small-$P_3$ linear extrapolation would overestimate its value at $P_s$ by about 130\%.
Together with the strong increase of $\tilde C^P_{3333}$ in Fig.~\ref{fig:fig1}(d), this evolution produces the large constant-$Q$ overestimate of $b_{333}$.
Near $P_s$, the positive slope of $\tilde a_{333}$ cannot account for the negative slope of $b_{333}$, showing that the rapid decrease of the compliance $(\tilde C^P_{3333})^{-1}$ dominates the turnover.

Microscopically, this stiffening is associated with the evolution of the Al-N bond network.
The layered hexagonal reference has a long axial Al-N separation, whereas the polar wurtzite state has a tetrahedrally coordinated Al-N network~\cite{Tsipas2013,Bacaksiz2015,Dreyer2016}.
Fig.~\ref{fig:fig2}(d) shows that increasing $P_3$ shortens the axial Al-N distance toward the in-plane Al-N bond-length.
As this separation approaches the in-plane bond length, axial deformation can no longer be accommodated easily by changing a long out-of-plane separation.
It increasingly involves stretching and bending the tetrahedrally connected Al-N bonds.
This provides a microscopic picture for the increase of $\tilde C^P_{3333}$ and for the changing mixed strain-polarization coupling $\tilde a_{333}$.

\textit{Comparison.---}
To separate this wurtzite behavior from a conventional ferroelectric reference, we apply the same analysis to PbTiO$_3$ using cubic $Pm\bar{3}m$ structure as the reference for tetragonal $P4mm$ phase~\cite{Haun1987,Yang2024,Zhao2025,yang2025ferroelectric}.

\begin{figure}[htbp]
    \centering
    \includegraphics[width=\columnwidth]{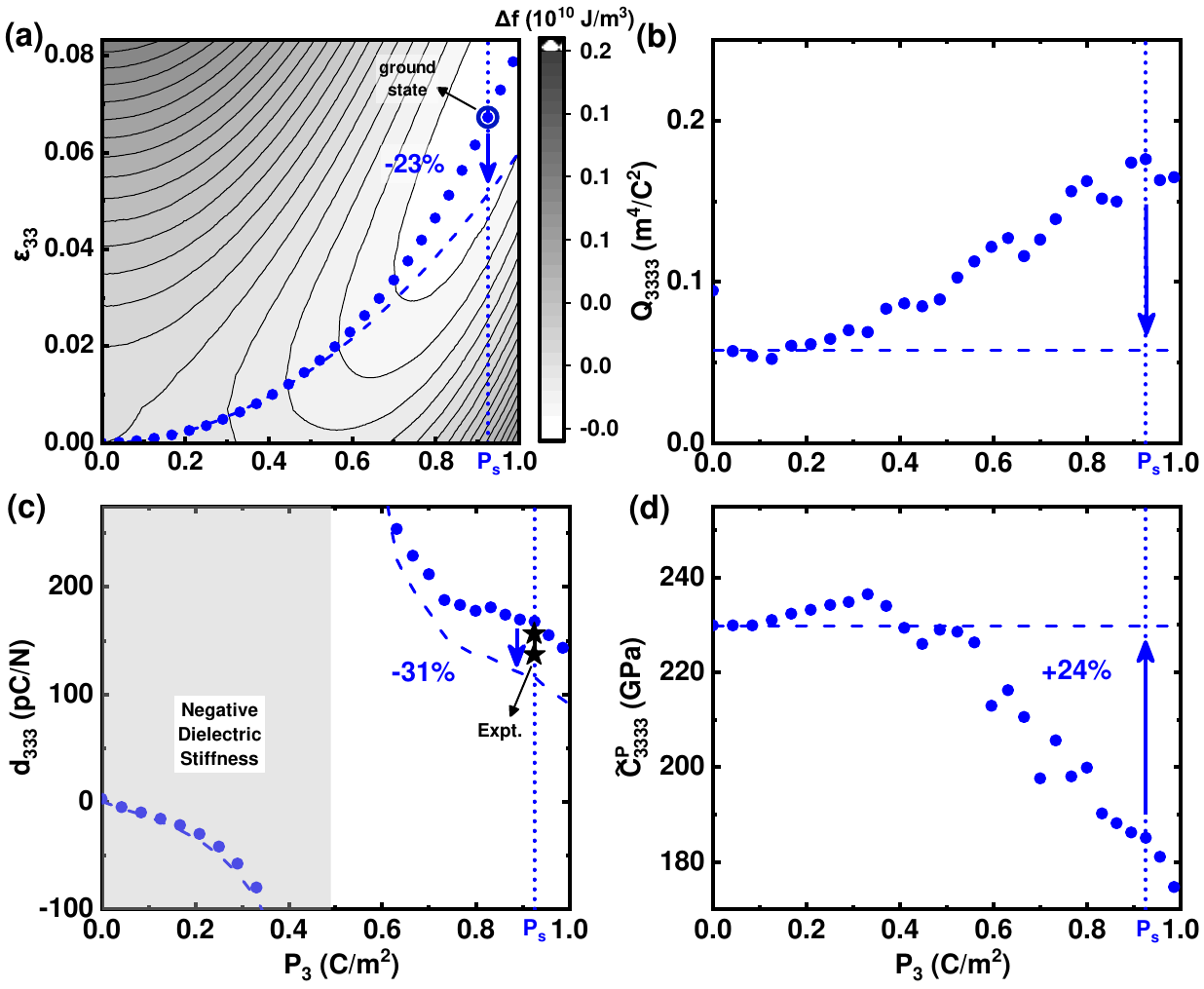}
    \caption{Polarization-dependent electromechanical response of PbTiO$_3$.
    (a) $\Delta f(P_3,\varepsilon_{33})$ with the DFT stress-free strain branch (blue dots) and its small-$P_3$ constant-$Q$ extrapolation (blue dashed), underestimating $\varepsilon^0_{33}(P_s)$ by 23\%.
    (b) $Q_{3333}$ and its $P_3=0$ reference.
    (c) $d_{333}$ and constant-$Q$ estimate, which underestimates it by 31\% at $P_s$. Two black stars mark two parameter sets fitted to Brillouin-scattering data for single-crystal in Ref.~\cite{Kalinichev1997}. Shading marks the negative-stiffness region with $(\partial E_3/\partial P_3)_\sigma<0$.
    (d) $\tilde C^P_{3333}$ and its $P_3=0$ reference, overestimating $C^P_{3333}(P_s)$ by 24\%. 
    Vertical dotted lines mark $P_s$.
}
\label{fig:fig3}
\end{figure}

In PbTiO$_3$, the stress-free strain remains much closer to a quadratic function of $P_3$ [Fig.~\ref{fig:fig3}(a)], $Q_{3333}$ varies more moderately [Fig.~\ref{fig:fig3}(b)], and the constant-$Q$ error in $d_{333}(P_s)$ is only 31\% [Fig.~\ref{fig:fig3}(c)].
The calculated $d_{333}(P_s)=165$~pC/N at $0$~K compares reasonably with the reported single-crystal range of $137$--$156$~pC/N~\cite{Kalinichev1997}.
The secant ratio $\varepsilon^0_{33}(P_s)/P_s^2=0.078$~m$^4$/C$^2$ is also close to the reported value of $0.08$~m$^4$/C$^2$~\cite{Gavrilyachenko1971,Adachi2001,Carpenter1998}.
Unlike AlN, the polar distortion in PbTiO$_3$ occurs within a connected perovskite framework, and $\tilde C^P_{3333}$ shows a weaker, opposite dependence, decreasing toward $P_s$ [Fig.~\ref{fig:fig3}(d)].

\begin{figure}[htbp]
    \centering
    \includegraphics[width=5.0cm]{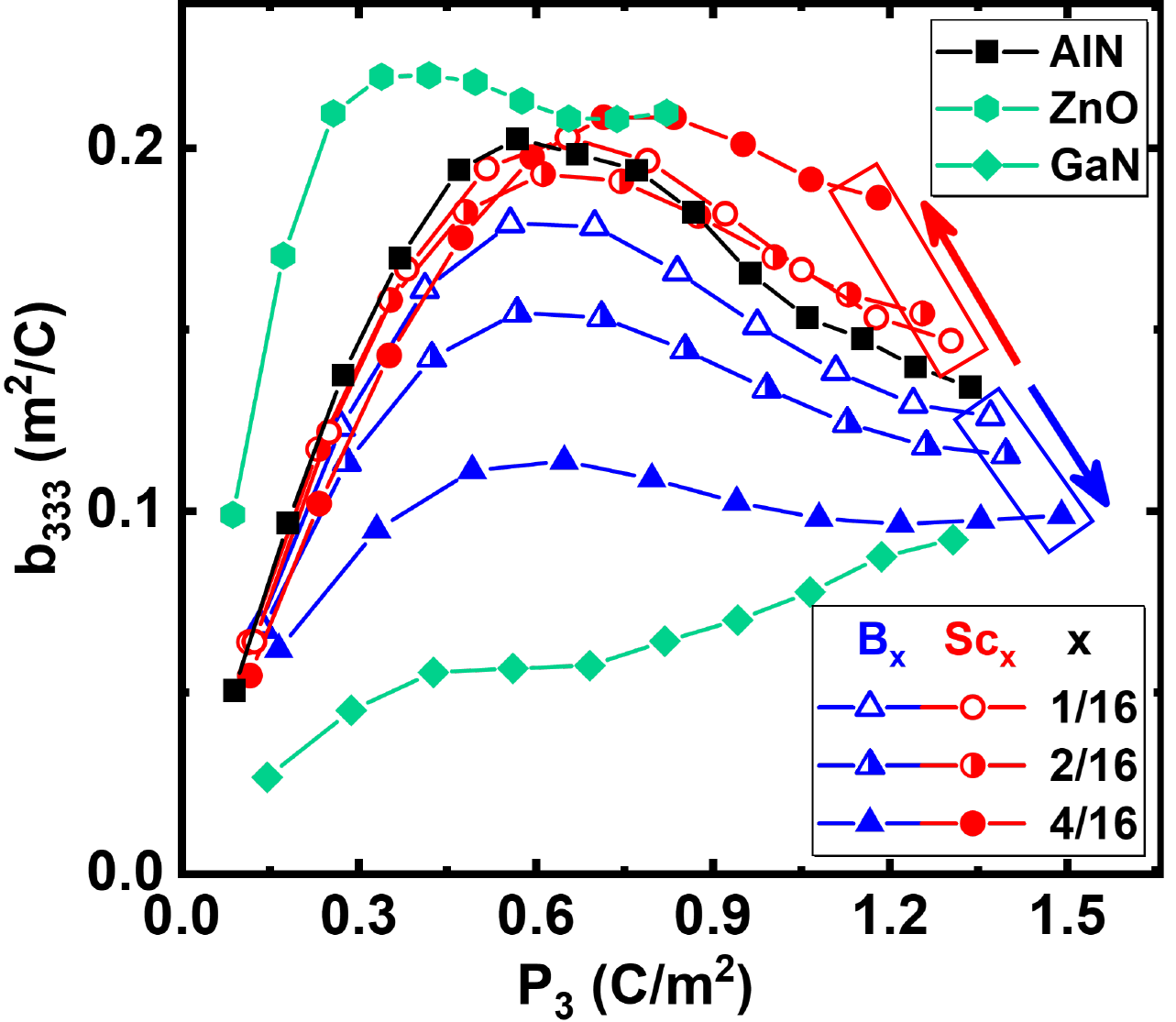}
    \caption{Polarization-dependent $b_{333}$ in ZnO, AlN, GaN, Al$_{1-x}$B$_x$N, and Al$_{1-x}$Sc$_x$N, showing deviations from the constant-$Q$ scaling $b_{333} \propto P_3$.
    Alloy configurations are from Ref.~\cite{Kanouni2023}. 
    Lines are guides to the eye.
    }\label{fig:fig4}
\end{figure}

\textit{Generality across wurtzites.---}
Fig.~\ref{fig:fig4} extends the fixed-stress strain-polarization factor $b_{333}$ to ZnO, GaN, and B- and Sc-substituted AlN using the same analysis.
All systems deviate from the constant-$Q$ scaling $b_{333}\propto P_3$, showing that polarization-dependent strain-polarization coupling is not unique to AlN.
Within the alloys, Sc substitution at $x=4/16$ enhances $b_{333}$ near $P_s$ while reducing $P_s$, consistent with coordination competition and elastic softening~\cite{Tasnadi2010}.
B substitution instead shifts the $b_{333}(P_3)$ curve downward while increasing $P_s$,
underscoring that Al$_{1-x}$B$_x$N follows a distinct bonding and structural route from Sc-substituted AlN~\cite{Mercado2026}.
It therefore does not show the same $b_{333}$ enhancement as Sc substitution.

\textit{Summary.---}
Our first-principles results show that, in strongly polar wurtzites, the large polarization strongly modifies the nature of electromechanical coupling beyond the quadratic polarization strain coupling.
Structurally, in AlN, the failure of constant electrostriction is associated with the evolution from a layered reference toward a tetrahedrally coordinated wurtzite network, which increases the reduced clamped-polarization modulus and changes the mixed strain-polarization coupling.
Thus, polarization-dependent electrostriction is not only a correction to continuum modeling, but also a design principle for engineering wurtzite piezoelectrics. 
Tuning the polarization-induced structural reconstruction can modify both the piezoelectric response and the elastic stiffness, offering possible routes to tailor electromechanical and acoustic properties.

\textit{Acknowledgment.---}
G.D.Z. thanks Y.J. Gu of Missouri S\&T for insightful discussions. 
This work was supported as part of the Computational Materials Sciences Program funded by the U.S. Department of Energy, Office of Science, Basic Energy Sciences, under Award No. DE-SC0020145. 
G.D.Z. and L.Q.C. also appreciate the generous support from the Donald W. Hamer Foundation through a Hamer Professorship at Penn State.
A.R. acknowledges the support of the National Science Foundation Graduate Research Fellowship Program under Grant No. DGE1255832.
Y.T. thanks the support from the National Science Foundation under grant DMR-2011839 through the MRSEC Center for Nanoscale Science of the Pennsylvania State University. 
J.-M. H. also acknowledges partial support for manuscript preparation from the Air Force Office of Scientific Research under award number FA9550-24-1-0159.

%

\end{document}


\preprint{APS/123-QED}

\title{\textbf{Supplementary Material for
``Anomalous Piezoelectricity from Polarization-Dependent Electrostriction in Wurtzites''}}

\author{Guo-Dong Zhao}
 \affiliation{
 Department of Materials Science and Engineering, Pennsylvania State University, University Park, Pennsylvania 16802, USA
}

\author{Aiden M. Ross}
 \affiliation{
 Department of Materials Science and Engineering, Pennsylvania State University, University Park, Pennsylvania 16802, USA
}

\author{Yueze Tan}
 \affiliation{
 Department of Materials Science and Engineering, Pennsylvania State University, University Park, Pennsylvania 16802, USA
}

\author{Yujie Zhu}
 \affiliation{
 Department of Materials Science and Engineering, University of Wisconsin-Madison, Madison, Wisconsin 53706, USA
}

\author{Jiamian Hu}
 \affiliation{
 Department of Materials Science and Engineering, University of Wisconsin-Madison, Madison, Wisconsin 53706, USA
}

\author{Long-Qing Chen}
 \email{lqc3@psu.edu}
 \affiliation{
 Department of Materials Science and Engineering, Pennsylvania State University, University Park, Pennsylvania 16802, USA
}

\date{\today}

\maketitle
\tableofcontents

\section{Thermodynamic framework and reduced coefficients}

\subsection{Notation and reduced free-energy landscape}
Throughout this Supplementary Material we use engineering-strain Voigt vectors
$\boldsymbol{\varepsilon}=(\varepsilon_{11},\varepsilon_{22},\varepsilon_{33},2\varepsilon_{23},2\varepsilon_{13},2\varepsilon_{12})^{\mathrm T}$
and $\allowbreak$
$\boldsymbol{\sigma}=(\sigma_{11},\sigma_{22},\sigma_{33},\sigma_{23},\sigma_{13},\sigma_{12})^{\mathrm T}$,
with bold symbols denoting Voigt vectors and matrices.
The polarization variable is the order-parameter polarization used in the main text.
For compactness we write it as $\mathbf P=(P_1,P_2,P_3)^{\mathrm T}$ in this Supplementary Material, with the understanding that it corresponds to $\mathbf P^s$ in the main text.
In the first-principles calculations it is represented by the zero-field Berry-phase polarization along the chosen polar-distortion path.

The main text identifies the zero-field first-principles surface with the $P^s$- and strain-dependent internal part of the chemo-mechanical free energy.
Before mechanical reduction, the derivative of the DFT surface with respect to the order-parameter polarization defines a thermodynamic conjugate field. 
After the reduction below, this becomes
$\mathscr E_3=(\partial f_{\rm eff}/\partial P_3)_{\varepsilon_3}$.
Along an equilibrium branch of the fixed-field potential $f_{\rm eff}-E_3P_3$, stationarity gives $\mathscr E_3=E_3$.
We keep the symbol $\mathscr{E}$ below so that the distinction between the internal derivative and the externally applied field remains explicit.
We also retain the subscript $P$ on susceptibilities, $\chi_P$, to emphasize that they refer to this order-parameter polarization coordinate.
For the reduced longitudinal response, $\varepsilon_0\chi^\sigma_{P,33}=(\partial P_3/\partial E_3)_\sigma$.

In this work we consider a uniaxial distortion path with $P_1=P_2=0$ and no homogeneous shear, 
$\varepsilon_4=\varepsilon_5=\varepsilon_6=0$.
The in-plane normal strains are symmetry related, $\varepsilon_1=\varepsilon_2\equiv\varepsilon_\perp$.
For each specified pair $(P_3,\varepsilon_3)$, the remaining internal coordinates and the in-plane strain are relaxed while enforcing $\sigma_1=\sigma_2=0$.
This procedure defines the reduced effective landscape
\begin{equation}
    f_{\rm eff}(P_3,\varepsilon_3) 
    \equiv 
    \min_{\varepsilon_\perp,\mathbf u} f_{\rm DFT}(P_3,\varepsilon_3,\varepsilon_\perp,\mathbf u), 
    \quad \text{with } \sigma_\perp=0.
    \label{eq:SM_projected_f}
\end{equation}
where $\mathbf u$ denotes all internal coordinates.
All reduced longitudinal response coefficients used in the main text are derivatives of this reduced surface and are therefore defined for the same transverse stress-free boundary condition.

\subsection{Eliminating relaxed strain variables under fixed-stress conditions}
We next show explicitly how fixed-stress mechanical boundary conditions enter the response coefficients.
The procedure is straightforward: 
when a stress component is held fixed, the conjugate strain component adjusts and can be eliminated from the constitutive equations.

The linearized constitutive relation around a given state may be written as
\begin{equation}
    \mathrm{d} \boldsymbol{\sigma}
    =
    \boldsymbol{C}^{P} \mathrm{d} \boldsymbol{\varepsilon}
    -\boldsymbol{a}\,\mathrm{d} \mathbf P,
\label{eq:SM_sigma_general_repeat}
\end{equation}
where
$\boldsymbol{C}^{P}\equiv(\partial^2 f_{\rm{DFT}}/\partial \boldsymbol{\varepsilon}\partial\boldsymbol{\varepsilon})_P$
is the clamped-order-parameter, or fixed-$P$, elastic stiffness matrix, and
$\boldsymbol{a}\equiv-(\partial^2 f_{\rm{DFT}}/\partial \boldsymbol{\varepsilon}\partial \boldsymbol{P})$
is the electromechanical stress coupling associated with the order-parameter polarization, both evaluated at the state about which the response is linearized.
We partition the strain and stress components into a relaxed block $r$, whose stresses are held fixed, and a retained (kept) block $k$, whose response remains explicit:
\begin{equation}
\begin{pmatrix}
\mathrm{d}\boldsymbol{\sigma}_r\\
\mathrm{d}\boldsymbol{\sigma}_k
\end{pmatrix}
=
\begin{pmatrix}
\boldsymbol C^P_{rr} & \boldsymbol C^P_{rk}\\
\boldsymbol C^P_{kr} & \boldsymbol C^P_{kk}
\end{pmatrix}
\begin{pmatrix}
\mathrm{d}\boldsymbol{\varepsilon}_r\\
\mathrm{d}\boldsymbol{\varepsilon}_k
\end{pmatrix}
-
\begin{pmatrix}
\boldsymbol a_r\\
\boldsymbol a_k
\end{pmatrix}\mathrm{d}\mathbf P. 
\label{eq:SM_block_constitutive}
\end{equation}
The fixed-stress condition on the relaxed block is $d\boldsymbol{\sigma}_r=0$.
The first row of Eq.~\eqref{eq:SM_block_constitutive} then gives
\begin{equation}
\mathrm{d}\boldsymbol{\varepsilon}_r
=
-(\boldsymbol C^P_{rr})^{-1}\boldsymbol C^P_{rk}\,\mathrm{d}\boldsymbol{\varepsilon}_k
+
(\boldsymbol C^P_{rr})^{-1}\boldsymbol a_r\,\mathrm{d}\mathbf P .
\label{eq:SM_eliminate_eps_a}
\end{equation}
Substituting Eq.~\eqref{eq:SM_eliminate_eps_a} into the second row of Eq.~\eqref{eq:SM_block_constitutive} gives the reduced constitutive relation
\begin{equation}
    \mathrm{d}\boldsymbol{\sigma}_k
    =
    \tilde{\boldsymbol C}^{P}_{kk}\,\mathrm{d}\boldsymbol{\varepsilon}_k
    -\tilde{\boldsymbol a}_{k}\,\mathrm{d}\mathbf P,
    \label{eq:SM_reduced_constitutive_general}
\end{equation}
where the reduced coefficients are
\begin{equation}
\tilde{\boldsymbol C}^{P}_{kk}
=
\boldsymbol C^{P}_{kk}
-\boldsymbol C^{P}_{kr}
(\boldsymbol C^{P}_{rr})^{-1}
\boldsymbol C^{P}_{rk},
\qquad
\tilde{\boldsymbol a}_{k}
=
\boldsymbol a_{k}
-\boldsymbol C^{P}_{kr}
(\boldsymbol C^{P}_{rr})^{-1}
\boldsymbol a_{r}.
\label{eq:SM_Schur_general}
\end{equation}
Thus \(\tilde{\boldsymbol C}^{P}_{kk}\) and \(\tilde{\boldsymbol a}_k\) are the elastic and electromechanical coefficients seen by the retained (kept) variables after the relaxed strain block has been eliminated under \(\mathrm{d}\boldsymbol{\sigma}_r=0\).
Mathematically, this block elimination gives the Schur-complement reduction of \(\boldsymbol C^P_{rr}\) in the full stiffness matrix~\cite{Horn2012}.
If the retained stresses are also held fixed, $\mathrm{d}\boldsymbol{\sigma}_k=0$, all relevant stresses are fixed and Eq.~\eqref{eq:SM_reduced_constitutive_general} gives
\begin{equation}
    \left(
    \frac{\partial\boldsymbol{\varepsilon}_k}{\partial\mathbf P}
    \right)_{\boldsymbol{\sigma}}
    =
    (\tilde{\boldsymbol C}^{P}_{kk})^{-1}\tilde{\boldsymbol a}_{k}.
\label{eq:SM_fixed_stress_branch_Q}
\end{equation}
In the present work the retained block contains only the longitudinal strain component, so $\tilde{\boldsymbol C}^{P}_{kk}$ is a $1\times1$ block.
Its inverse is the reduced longitudinal compliance used in the main text,
\[
S^P_{3333}=(\tilde{C}^P_{3333})^{-1},
\]
used for the reduced one-dimensional branch in the main text.

\subsection{One-dimensional response identities after mechanical reduction}
After the transverse strains have been eliminated, the main-text analysis uses the one-dimensional reduced surface $f_{\rm eff}(P,\varepsilon)$, where $P\equiv P_3$ and $\varepsilon\equiv\varepsilon_3$.
All scalar coefficients in this subsection are reduced coefficients derived from $f_{\rm eff}$, therefore tildes are omitted for compactness.
The conjugate variables are
\begin{equation}
\mathscr{E}=\left(\frac{\partial f_{\rm eff}}{\partial P}\right)_{\varepsilon},
\qquad
\sigma=\left(\frac{\partial f_{\rm eff}}{\partial \varepsilon}\right)_{P},
\end{equation}
where $\mathscr{E}$ is the thermodynamic field conjugate to the order-parameter polarization. 
Along the equilibrium branch of the fixed-field potential $f_{\rm eff}-E P$, stationarity gives $\mathscr E=E$, so variations satisfy $\mathrm{d}\mathscr E=\mathrm{d}E$.
Linearizing around a given state gives
\begin{equation}
\begin{pmatrix}
\mathrm{d}\sigma\\
\mathrm{d}\mathscr{E}
\end{pmatrix}
=
\begin{pmatrix}
f_{\varepsilon\varepsilon} & f_{\varepsilon P}\\
f_{P\varepsilon} & f_{PP}
\end{pmatrix}
\begin{pmatrix}
\mathrm{d}\varepsilon\\
\mathrm{d}P
\end{pmatrix},
\label{eq:SM_Hessian_1D}
\end{equation}
where all derivatives are evaluated on the smooth fitted surface $f_{\rm eff}(P,\varepsilon)$ at the point of interest. 
Thus the Hessian of the reduced surface $f_{\rm eff}(P,\varepsilon)$ directly provides the fixed-variable coefficients $C^P$, $\chi^\varepsilon_P$, and $a$, while the fixed-stress coefficients $b$ and $\chi^\sigma_P$ are obtained only after imposing $\mathrm d\sigma=0$.
With the sign convention used in the main text,
\[
C^P=f_{\varepsilon\varepsilon},\qquad
a=-f_{\varepsilon P}=-f_{P\varepsilon}.
\]
Thus the two linearized scalar equations are
\begin{align}
\mathrm{d} \sigma      &=  C^P \mathrm{d}\varepsilon-a\,\mathrm{d}P,  \label{eq:SM_1D_dsigma}\\
\mathrm{d} \mathscr{E} &= -a\, \mathrm{d}\varepsilon+f_{PP} \mathrm{d}P. \label{eq:SM_1D_dE}
\end{align}

Under fixed stress, $\mathrm{d}\sigma=0$, 
Eq.~\eqref{eq:SM_1D_dsigma} then gives
\[
\mathrm{d}\varepsilon=\frac{a}{C^P} \mathrm{d}P.
\]
Therefore
\begin{equation}
b\equiv
\left(\frac{\partial\varepsilon}{\partial P}\right)_\sigma
=
\frac{a}{C^P}
=
S^P a.
\label{eq:SM_deps_dP_sigma}
\end{equation}
Substituting $\mathrm{d}\varepsilon=(a/C^P)\mathrm{d}P$ into Eq.~\eqref{eq:SM_1D_dE} gives
$\mathrm{d}\mathscr{E}=
    \left(
    f_{PP}-\frac{a^2}{C^P}
    \right)\mathrm{d}P$.
Therefore the fixed-stress order-parameter susceptibility is
\begin{equation}
\left(\frac{\partial P}{\partial \mathscr{E}}\right)_\sigma
=
\left[
f_{PP}-\frac{a^2}{C^P}
\right]^{-1},
\qquad
\chi^\sigma_P=\varepsilon_0^{-1}\left(\frac{\partial P}{\partial \mathscr{E}}\right)_\sigma.
\label{eq:SM_chi_sigma}
\end{equation}
Here $\chi^\sigma_P$ is the dimensionless order-parameter susceptibility defined in Eq.~\eqref{eq:SM_chi_sigma}.
At equilibrium $\mathscr{E}=E$, so the order-parameter-mediated converse piezoelectric coefficient is
\begin{equation}
d=
\left(\frac{\partial\varepsilon}{\partial E}\right)_\sigma
=
\left(\frac{\partial\varepsilon}{\partial \mathscr{E}}\right)_\sigma
=
\left(\frac{\partial\varepsilon}{\partial P}\right)_\sigma
\left(\frac{\partial P}{\partial \mathscr{E}}\right)_\sigma
=
b\,\varepsilon_0\chi^\sigma_P.
\label{eq:SM_d_final}
\end{equation}

For completeness, the fixed-strain order-parameter susceptibility is
$\chi^\varepsilon_P=\varepsilon_0^{-1}(f_{PP})^{-1}$.
The fixed-field response of \(P\) to strain is obtained by holding the applied field constant.
Since \(\mathscr{E}=E\) along the equilibrium branch, fixed \(E\) implies \(d\mathscr{E}=dE=0\).
From Eq.~\eqref{eq:SM_1D_dE}, one obtains the corresponding reduced one-dimensional piezoelectric stress coefficient
\begin{equation}
e\equiv
\left(\frac{\partial P}{\partial\varepsilon}\right)_E
=
\frac{a}{f_{PP}},
\label{eq:SM_e_def}
\end{equation}
and
\begin{equation}
C^{E}\equiv
\left(\frac{\partial\sigma}{\partial\varepsilon}\right)_E
= C^P-\frac{a^2}{f_{PP}}.
\label{eq:SM_CE_def}
\end{equation}
Combining these definitions gives the one-dimensional fixed-stress/fixed-strain susceptibility relation
\begin{equation}
\chi^\sigma_P
=
\chi^\varepsilon_P+\frac{e^2}{\varepsilon_0 C^E}.
\label{eq:SM_chi_relation}
\end{equation}

\subsection{Transversely stress-free reduction for hexagonal/tetragonal symmetry}
The preceding subsection gives the general block-matrix reduction for eliminating mechanically relaxed strain components. 
For the hexagonal or tetragonal normal-strain sector used in the main text, the corresponding reduction can be written explicitly.
To illustrate the transverse-stress reduction analytically, consider a constant-coefficient normal-strain sector for polarization along $3$,
\begin{align}
\sigma_\perp &= (C^{P}_{11}+C^{P}_{12})\,\varepsilon_\perp
+ C^{P}_{13}\,\varepsilon_{3}
- q_{13}\,P_3^2,
\label{eq:SM_sigma_perp}\\
\sigma_{3} &= 2C^{P}_{13}\,\varepsilon_\perp
+ C^{P}_{33}\,\varepsilon_{3}
- q_{33}\,P_3^2,
\label{eq:SM_sigma_33}
\end{align}
where $\sigma_\perp$ denotes the common in-plane stress, $\sigma_\perp\equiv\sigma_1=\sigma_2$, and
$q_{i3}=-(\partial\sigma_i/\partial P_3^2)_{\varepsilon}$.
Note that Voigt notation is used in this subsection, so $\tilde C^P_{33}$ corresponds to $\tilde C^P_{3333}$ in tensor notation.
Imposing $\sigma_\perp=0$ gives 
$\varepsilon_\perp= 
    \frac{q_{13}P_3^2-C^{P}_{13}\varepsilon_3}
    {C^{P}_{11}+C^{P}_{12}}$.
Substitution into the expression for $\sigma_3$ gives the reduced longitudinal constitutive relation
\begin{equation}
\sigma_3=\tilde C^P_{33}\,\varepsilon_3-\tilde q_{33}\,P_3^2,
\label{eq:SM_effective_1D}
\end{equation}
with
\begin{align}
\tilde C^P_{33}
&=
C^{P}_{33}
-\frac{2(C^{P}_{13})^2}{C^{P}_{11}+C^{P}_{12}},
\label{eq:SM_Ctilde}\\
\tilde q_{33}
&=
q_{33}
-\frac{2C^{P}_{13}q_{13}}{C^{P}_{11}+C^{P}_{12}}.
\label{eq:SM_qtilde}
\end{align}
Thus even in this lowest-order illustration, the longitudinal stress-free strain response depends on reduced combinations $\tilde C_{33}$ and $\tilde q_{33}$, rather than on the unreduced ratio $q_{33}/C^P_{33}$.
This explicitly shows why the transverse mechanical response must be eliminated under the $\sigma_\perp=0$ before defining the reduced one-dimensional coefficients $b_{333}$, $\tilde C^P_{3333}$, and $\tilde a_{333}$ used in the main text.

\section{First-principles computational details and validation}

\subsection{Density functional theory methods}
All first-principles calculations were performed using density functional theory (DFT) with the projector-augmented-wave (PAW) method implemented in Vienna \textit{Ab initio} Simulation Package (\textsc{vasp})~\cite{Kresse1996-1,Kresse1996-2}.
Unless stated otherwise, we used a plane-wave cutoff of 600~eV and the revised Perdew-Burke-Ernzerhof for solids (PBEsol) exchange-correlation functional~\cite{Perdew2008} within the generalized gradient approximation.
The valence electrons considered in pseudopotentials are 3 for Al, 5 for N, 14 for Pb, 12 for Ti, and 6 for O.
The Brillouin zone was sampled with a maximum $k$-point spacing below $0.2$~\AA$^{-1}$ (e.g., $12\times 12\times 8$ for AlN and $9\times 9\times 9$ for PbTiO$_3$).
The electronic energy convergence threshold was $1~\mu$eV, and structural relaxations were converged until all residual forces were below $2$~meV/\AA.
For the free-energy grid points used to take numerical derivatives, we used a tighter electronic tolerance of 0.1~$\mu$eV to reduce numerical noise.
The data shown in Fig.~4 of the main text were obtained from fixed polar-mode amplitude calculations under stress-free conditions. 
These calculations reproduce the bold longitudinal stress-free branch used to evaluate $b_{333}=(\partial\varepsilon_{33}/\partial P_3)_\sigma$ and are therefore sufficient for comparing the strain-polarization factor across a broader set of wurtzite materials.

\begin{figure}[hbtp]
\centering
\includegraphics[width= 10cm]{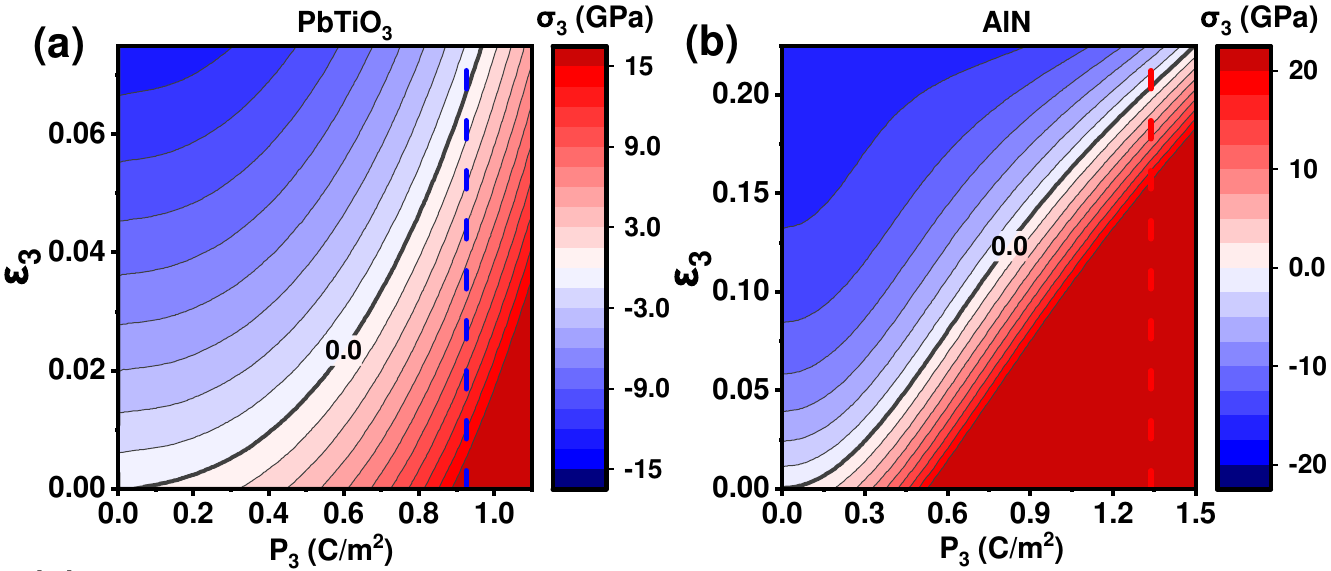}
    \caption{First-principles stress contours $\sigma_{3}(P_{3},\varepsilon_{3})$ after relaxing the remaining degrees of freedom under $\sigma_\perp=0$ for 
    (a) PbTiO$_3$ and (b) AlN.
    The thick curve denotes $\sigma_{3}=0$.
    The blue and red dashed lines indicate the spontaneous polarizations of PbTiO$_3$ and AlN.
    The characteristic near-parabolic strain-polarization behavior in PbTiO$_3$ and the pronounced deviation from it in AlN persist under finite longitudinal stress.
    }\label{fig:figs1}
\end{figure}

\begin{table}[hbp]
\caption{\label{tab:table1}%
    Calculated $c$-lattice constants for the layered hexagonal ($c_h$) and wurtzite ($c_w$) phases of AlN without and with different types of van der Waals corrections:
    optB86b-vdW (optb86b), DFT-D3 method with Becke-Johnson damping function (D3), and Tkatchenko-Scheffler method with iterative Hirshfeld partitioning (TS)
    }
\centering
\begin{tabular*}{0.75\textwidth}{@{\extracolsep{\fill}}lccc}
\toprule\toprule
vdW & $c_h$ (\AA) & $c_w$ (\AA) & $\varepsilon_{3}(P_s)$ \\
\midrule
w/o vdW & 4.186 & 5.015 & 0.198 \\
optb86b & 4.169 & 5.002 & 0.200 \\
D3      & 4.140 & 4.975 & 0.202 \\
TS      & 4.166 & 5.002 & 0.201 \\
\bottomrule\bottomrule
\end{tabular*}
\end{table}

\begin{figure}[hbtp]
\centering
\includegraphics[width= 10cm]{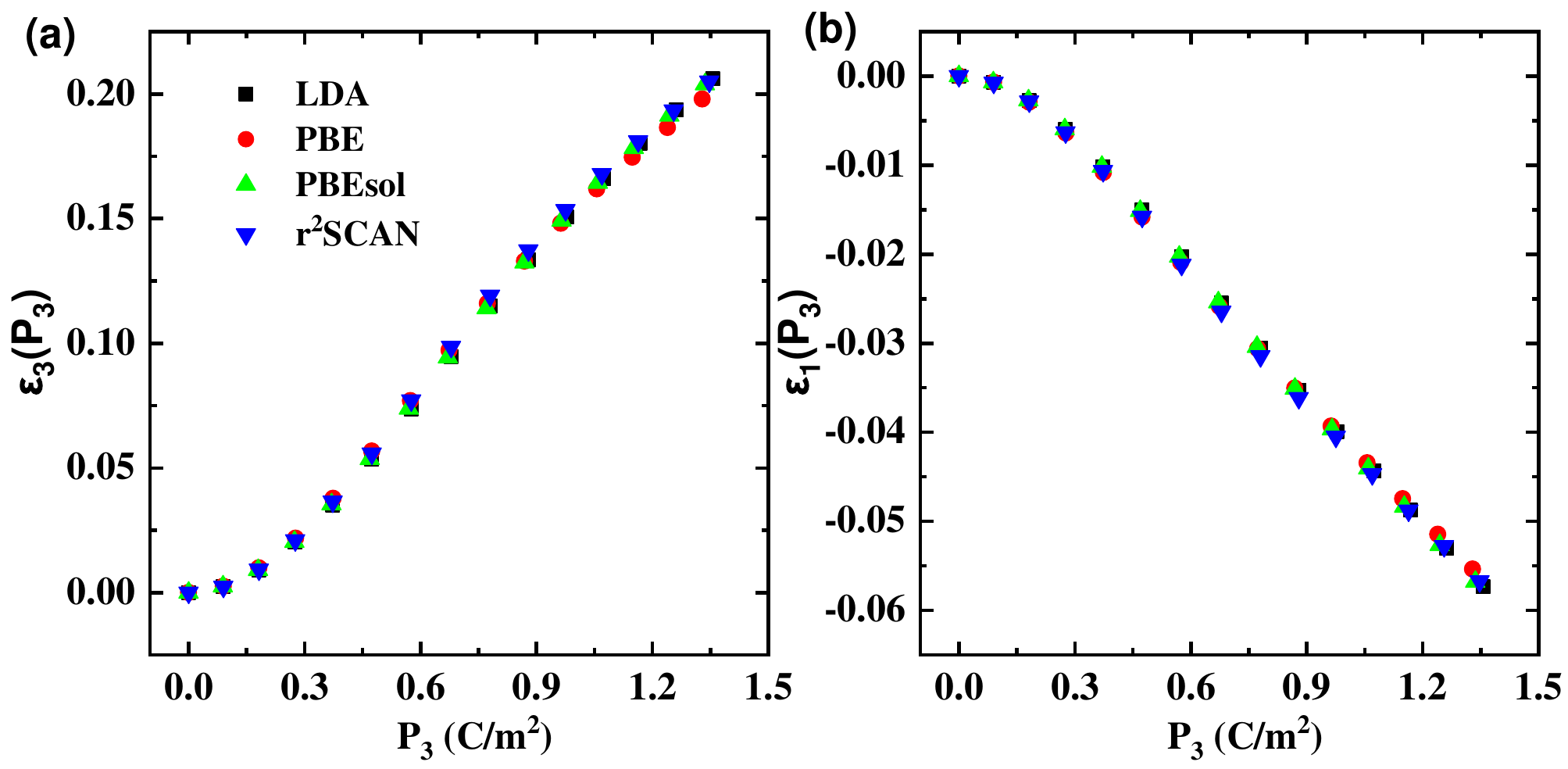}
    \caption{Calculated spontaneous strain branches (a) $\varepsilon_{3}(P_3)$ and (b) $\varepsilon_{1}(P_3)$ for AlN under stress-free conditions, using different exchange-correlation functionals.
    }\label{fig:figs2}
\end{figure}

Fig.~\ref{fig:figs2} and Table~\ref{tab:table1} show that the calculated spontaneous strain branches and the resulting conclusions are robust with respect to the choice of exchange-correlation functional and van der Waals correction.

\subsection{Constructing and postprocessing the reduced first-principles free-energy surface}
For each material, we first determined the ferroelectric (polar) ground state.
To construct the free-energy surfaces used in the main text, we generated a two-dimensional grid of initial structures by interpolating between the reference nonpolar structure and the relaxed polar ground state, using a fractional step size of 0.04 in both the polar-mode amplitude and the homogeneous axial strain.
At each grid point we fixed the axial strain $\varepsilon_{3}$ and the polar-mode amplitude, which serves as the practical constraint variable for polar distortion.
The remaining internal coordinates and the in-plane strains were then relaxed subject to $\sigma_\perp=0$.
The corresponding $P_3$ value was obtained from the zero-field Berry-phase polarization following Ref.~\cite{King1993}.
The resulting energies and stresses provide a discrete sampling of the reduced first-principles surface $f_{\mathrm{eff}}(P_3,\varepsilon_{3})$ defined in Eq.~\eqref{eq:SM_projected_f}.

For numerical differentiation, we use a reference-frame convention.
The cell dipole moment is expressed as a polarization density per reference volume,
\begin{equation}
    P_3^{\rm ref}
    =
    \frac{V}{V_{\rm ref}}P_3^{\rm raw},
    \qquad
    f_{\rm ref}
    =
    \frac{U_{\rm cell}}{V_{\rm ref}},
    \label{eq:SM_ref_frame}
\end{equation}
where $P_3^{\rm raw}$ is the Berry-phase polarization normalized by the instantaneous cell volume $V$, $U_{\rm cell}$ is the DFT total energy of the cell, and $V_{\rm ref}$ is the volume of the chosen reference structure.
This convention keeps polarization, strain, and free-energy density in the same Lagrangian description.
Derivatives such as
\begin{equation}
    \mathscr E_3
    =
    \left(
    \frac{\partial f_{\rm ref}}
         {\partial P_3^{\rm ref}}
    \right)_{\varepsilon_3},
    \qquad
    \sigma_3
    =
    \left(
    \frac{\partial f_{\rm ref}}
         {\partial \varepsilon_3}
    \right)_{P_3^{\rm ref}}
    \label{eq:SM_ref_conjugate}
\end{equation}
are evaluated with the stress sign and volume convention consistent with Eq.~\eqref{eq:SM_ref_conjugate}.

The stress-free branch is obtained from $\sigma_3=0$ along fixed polar-mode-amplitude scan lines.
In practice, we linearly interpolate between adjacent axial-strain points whose calculated stresses bracket zero and use the lowest-energy crossing if multiple crossings occur.
This gives a discrete stress-free branch $\varepsilon^0_{33}(P_3^{\rm ref})$.
Because the Berry-phase polarization is not uniformly spaced after structural relaxation and the reference-frame transformation in Eq.~\eqref{eq:SM_ref_frame}, all derivatives are evaluated from the actual DFT coordinates in the $(P_3^{\rm ref},\varepsilon_3)$ plane rather than from the nominal grid indices.

Local derivatives on the stress-free branch are obtained from small two-dimensional least-squares fits to neighboring grid points.
Near $P_3=0$, where only nonnegative polarizations are sampled, we use
\begin{equation}
    u=(P_3^{\rm ref})^2
\end{equation}
to enforce the even symmetry of the energy and stress with respect to $P_3$.
Writing
\begin{equation}
    \Delta u=u-u_0,
    \qquad
    \Delta\varepsilon_3=\varepsilon_3-\varepsilon^0_3,
\end{equation}
we fit the local stress surface as
\begin{equation}
    \sigma_3(u,\varepsilon_3)
    \simeq
    s_0+s_u\Delta u+s_\varepsilon\Delta\varepsilon_3 ,
    \label{eq:SM_local_sigma_fit}
\end{equation}
and the local energy surface as
\begin{align}
    f_{\rm ref}(u,\varepsilon_3)
    \simeq\;&
    c_0+c_u\Delta u+c_\varepsilon\Delta\varepsilon_3
    +\frac{1}{2}c_{uu}\Delta u^2
    +c_{u\varepsilon}\Delta u\,\Delta\varepsilon_3
    +\frac{1}{2}c_{\varepsilon\varepsilon}\Delta\varepsilon_3^2 .
    \label{eq:SM_local_energy_fit}
\end{align}
The stress fit gives
\begin{equation}
    f_{\varepsilon\varepsilon}=s_\varepsilon,
    \qquad
    f_{P\varepsilon}
    =
    \frac{\partial\sigma_3}{\partial P_3^{\rm ref}}
    =
    2P_3^{\rm ref}s_u ,
    \label{eq:SM_fPe_from_sigma}
\end{equation}
so that the symmetry-required limit $f_{P\varepsilon}\to0$ as $P_3\to0$ is satisfied.
The energy fit gives
\begin{equation}
    f_{PP}
    =
    \frac{\partial^2 f_{\rm ref}}
         {\partial (P_3^{\rm ref})^2}
    =
    2c_u+4(P_3^{\rm ref})^2c_{uu}.
    \label{eq:SM_fPP_from_u}
\end{equation}
The corresponding $f_{P\varepsilon}$ and $f_{\varepsilon\varepsilon}$ obtained from the energy fit are used as internal consistency checks.

The fixed-stress dielectric stiffness is obtained from the Schur complement of the local Hessian,
\begin{equation}
    \left(\varepsilon_0\chi_{33}^{\sigma,\rm ref}\right)^{-1}
    =
    \left(
    \frac{\partial \mathscr E_3}
         {\partial P_3^{\rm ref}}
    \right)_{\sigma_3}
    =
    f_{PP}
    -
    \frac{f_{P\varepsilon}^2}{f_{\varepsilon\varepsilon}} .
    \label{eq:SM_invchi_method}
\end{equation}
The strain-polarization factor and the order-parameter-mediated converse piezoelectric response are then
\begin{align}
    b_{333}^{\rm ref}
    &=
    \left(
    \frac{\partial\varepsilon_{33}}
         {\partial P_3^{\rm ref}}
    \right)_{\sigma_3}
    =
    -\frac{f_{P\varepsilon}}{f_{\varepsilon\varepsilon}},
    \nonumber\\
    d_{333}
    &=
    \left(
    \frac{\partial\varepsilon_{33}}
         {\partial \mathscr E_3}
    \right)_{\sigma_3}
    =
    b_{333}^{\rm ref}\varepsilon_0\chi_{33}^{\sigma,\rm ref}.
    \label{eq:SM_d_method}
\end{align}
The curvature-defined electrostrictive coefficient reported in the main text is
\begin{equation}
    Q_{3333}^{\rm ref}(P_3)
    =
    \frac{1}{2}
    \frac{\mathrm d^2\varepsilon^0_{33}}
         {\mathrm d(P_3^{\rm ref})^2}
    =
    \frac{1}{2}
    \frac{\mathrm d b_{333}^{\rm ref}}
         {\mathrm dP_3^{\rm ref}},
    \label{eq:SM_Q_method}
\end{equation}
where the second equality holds on the exact stress-free branch.
This curvature-defined coefficient is distinct from the secant ratio $\varepsilon^0_{33}/P_3^2$; the two coincide only for a purely quadratic stress-free strain branch.

When results are displayed as functions of the raw current-volume polarization, derivative-defined coefficients are transformed by the chain rule.
With
\begin{equation}
    y_1=
    \frac{\mathrm dP_3^{\rm ref}}
         {\mathrm dP_3^{\rm raw}},
    \qquad
    y_2=
    \frac{\mathrm d^2P_3^{\rm ref}}
         {\mathrm d(P_3^{\rm raw})^2},
    \label{eq:SM_y_def}
\end{equation}
the raw-coordinate curvature is
\begin{equation}
    \frac{\mathrm d^2\varepsilon^0_{33}}
         {\mathrm d(P_3^{\rm raw})^2}
    =
    \frac{\mathrm d^2\varepsilon^0_{33}}
         {\mathrm d(P_3^{\rm ref})^2}y_1^2
    +
    \frac{\mathrm d\varepsilon^0_{33}}
         {\mathrm dP_3^{\rm ref}}y_2,
    \label{eq:SM_raw_Q_chain}
\end{equation}
and $Q_{3333}^{\rm raw}$ is one half of this value.
Similarly,
\begin{equation}
    b_{333}^{\rm raw}
    =
    b_{333}^{\rm ref}y_1,
    \qquad
    \chi_{33}^{\sigma,\rm raw}
    =
    \frac{\chi_{33}^{\sigma,\rm ref}}{y_1},
    \label{eq:SM_raw_b_chi_chain}
\end{equation}
so the product $b_{333}\varepsilon_0\chi_{33}^{\sigma}$ gives the same physical $d_{333}$ when the transformation is applied consistently.
The local fits are used only for numerical differentiation: no low-order global Landau polynomial or constant electrostrictive coefficient is imposed on the full polarization range.
The present work focuses on the reduced longitudinal coefficients obtained directly from the two-dimensional first-principles surface, while a full tensorial extension would require separate low-symmetry constrained-polarization perturbations.

\subsection{The bare displacement-based piezoelectric coefficient $g_{33}^{\rm bare}$}

\begin{figure}[hbtp]
\centering
\includegraphics[width= 16cm]{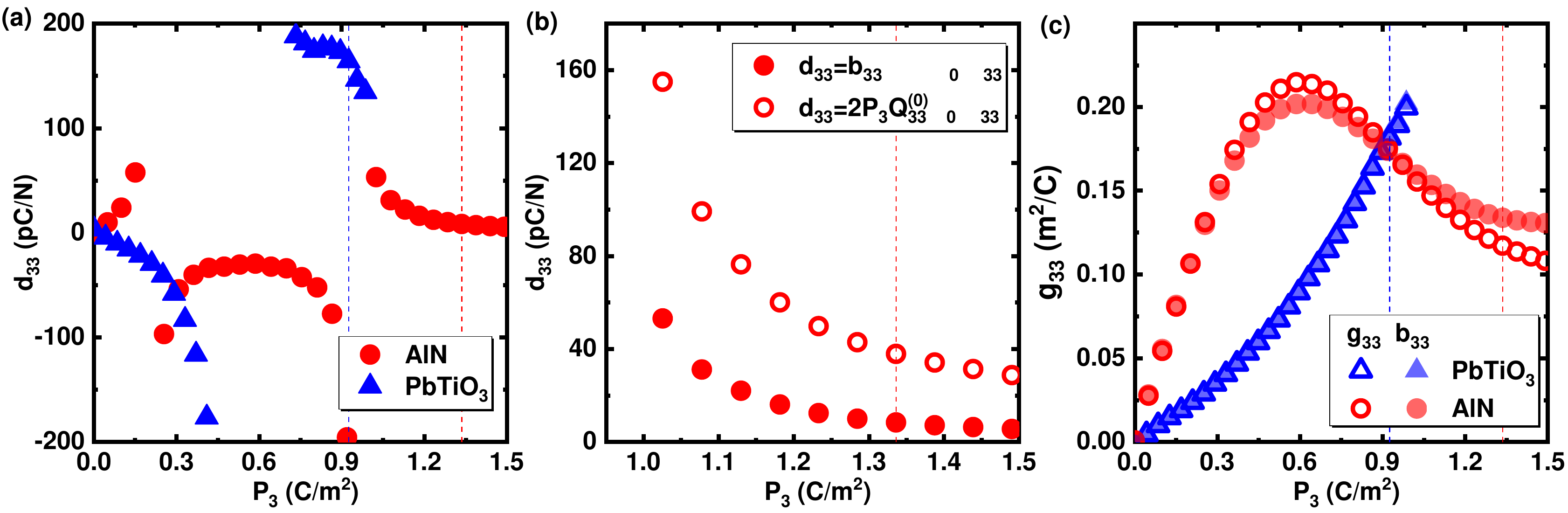}
    \caption{(a) Piezoelectric strain coefficient $d_{33}$ of PbTiO$_3$ (blue triangles) and AlN (red dots).
    (b) Comparison for AlN between the constant-$Q$ estimate of $d_{33}$ (red circles) and the value calculated from first-principles free-energy surface (red dots).
    (c) Bare displacement-based coefficient $g_{33}^{\rm bare}=(\partial \varepsilon_3/\partial D_3^{\rm bare})_\sigma$ obtained from the stress-free relation $\varepsilon_3(D_3^{\rm bare})$ for PbTiO$_3$ (blue triangles) and AlN (red dots).
    Blue and red vertical dashed lines mark the spontaneous polarizations of PbTiO$_3$ and AlN, respectively.
    }\label{fig:figs3}
\end{figure}

For completeness, we also report a displacement-based piezoelectric coefficient defined with a bare displacement variable, which excludes field-induced polarization components not included in the order-parameter coordinate~\cite{Tagantsev1986},
\begin{align}
    g_{33}^{\rm{bare}}=\left(\frac{\partial \varepsilon_{3}}{\partial D_3^{\rm{bare}}}\right)_\sigma,
    \label{eq:SM_g}
\end{align}
which provides a $D$-variable counterpart of the polarization-based decomposition used in the main text.

Using
\begin{align}
    D_3^{\rm{bare}}(P_3)=P_3+\varepsilon_0 \mathscr{E}_3(P_3),
    \label{eq:SM_D}
\end{align}
we obtain
\begin{align}
    g^{\rm{bare}}(P)=\left(\frac{\partial \varepsilon}{\partial D^{\rm{bare}}}\right)_\sigma
        =\left(\frac{\partial \varepsilon}{\partial P}\right)_\sigma
         \left(\frac{\partial P          }{\partial D^{\rm{bare}}}\right)_\sigma
        =\frac{b}
         {1+\big(\chi^\sigma_P\big)^{-1}}.
    \label{eq:SM_gP}
\end{align}
The corresponding relation to the converse piezoelectric coefficient mediated by the bare displacement response is
\begin{align}
    d=g^{\rm{bare}}\,\varepsilon_0\varepsilon_{r,\rm{bare}}^\sigma,
    \qquad
    \varepsilon_0\varepsilon_{r,\rm bare}^\sigma
    =\left(\frac{\partial D^{\rm{bare}}}{\partial E}\right)_\sigma
    =\varepsilon_0+\left(\frac{\partial P}{\partial E}\right)_\sigma.
    \label{eq:SM_dg}
\end{align}
Here $P$ is the order-parameter polarization used in the DFT free-energy surface.
The superscript "bare" means that $D^{\rm bare}=P+\varepsilon_0E$ contains only the order-parameter polarization and the vacuum displacement, and does not include additional field-induced polarization components outside order-parameter coordinate.
Such additional contributions are collected in the field-only term $\phi(E)$ in the thermodynamic potential~\cite{Tagantsev1986}.

\subsection{The stability of reference phases in AlN and PbTiO$_3$}

\begin{figure}[hbtp]
\centering
\includegraphics[width=12cm]{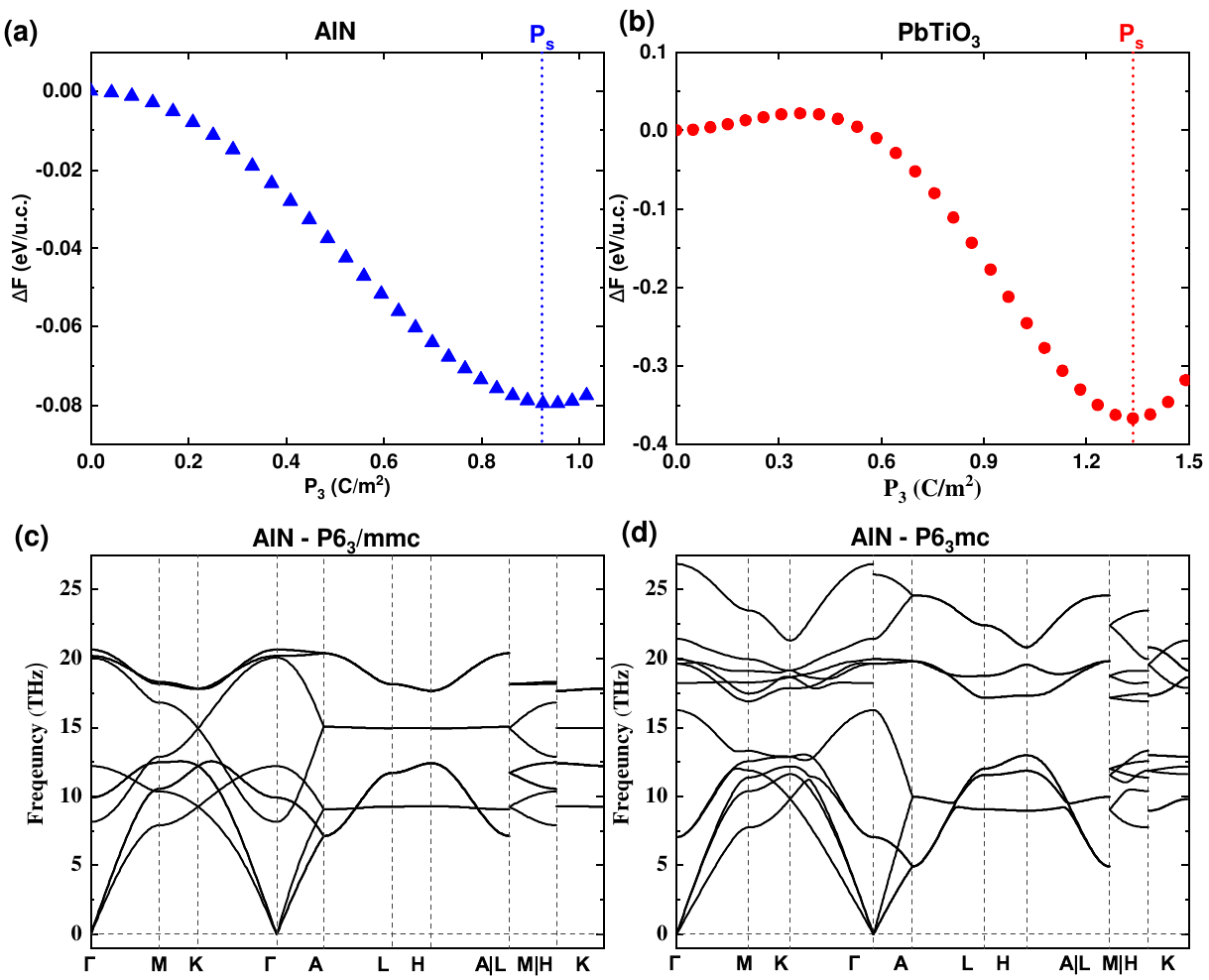}
    \caption{Stability of the reference phases in PbTiO$_3$ and AlN.
    DFT energy difference with respect to the reference phase along the stress-free branch of (a) PbTiO$_3$ and (b) AlN.
    Phonon dispersions of AlN in the (c) layered-hexagonal $P6_3/mmc$ reference phase and (d) polar wurtzite $P6_3mc$ phase.
    Non-analytical corrections are included in (d) to account for LO-TO splitting.
    }
    \label{fig:figs4}
\end{figure}

Fig.~\ref{fig:figs4} compares the stability of the high-symmetry reference structures used for PbTiO$_3$ and AlN.
For PbTiO$_3$, the cubic $Pm\bar{3}m$ reference is unstable at $0$~K along the stress-free ferroelectric distortion path, as shown by the negative curvature of the DFT energy away from $P_3=0$ [Fig.~\ref{fig:figs4}(a)].

On the other hand, the layered-hexagonal $P6_3/mmc$ reference structure of AlN is a local minimum on the stress-free DFT energy surface [Fig.~\ref{fig:figs4}(b)].
The phonon dispersion in Fig.~\ref{fig:figs4}(c) further shows that there are no imaginary phonon frequencies, indicating that this reference phase is dynamically stable within the present calculation.
Therefore, the nonpolar reference phase of AlN is a metastable structure separated from the ground polar wurtzite phase by a finite energy barrier.
The polar wurtzite $P6_3mc$ phase is also dynamically stable, as confirmed by the phonon dispersion in Fig.~\ref{fig:figs4}(d).

This distinction is important for interpreting the polarization-dependent electrostriction in the main text.
In PbTiO$_3$, the polar distortion develops from an unstable centrosymmetric perovskite reference.
In AlN, however, the polarization path connects two locally stable bonding environments, from a layered nonpolar reference to a tetrahedrally coordinated wurtzite network.
The large structural reorganization along this path provides the microscopic basis for the strong polarization dependence of $Q_{3333}$, $b_{333}$, and $\tilde C^P_{3333}$ discussed in the main text.

\section{Relation to Landau-type representations}

\subsection{Polarization-dependent elasticity in a local branch expansion}

\begin{figure}[hbtp]
\centering
\includegraphics[width= 14cm]{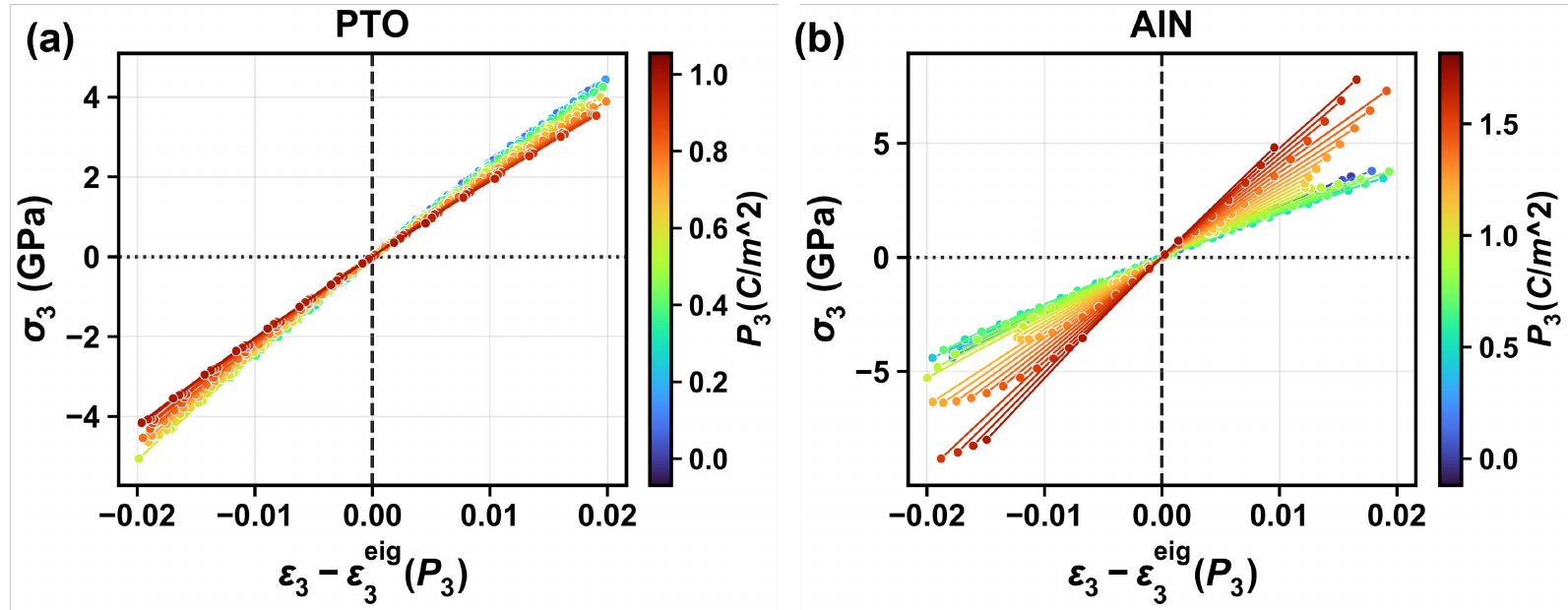}
    \caption{Elastic stress - elastic strain curves at different fixed-$P_3$ in (a) PbTiO$_3$ and (b) AlN.
    }\label{fig:figs5}
\end{figure}

Landau-type expansions provide a useful compact language for electromechanical coupling. 
Here we use a closely related local form around the stress-free branch of the first-principles surface.
Let the stress-free strain $\varepsilon^0(P)$ be defined by
\begin{equation}
\left.
\frac{\partial f_{\rm eff}}{\partial \varepsilon}
\right|_{P,\varepsilon^0(P)}
=0,
\label{eq:SM_sigma0_branch}
\end{equation}
and define the elastic strain relative to this branch,
\begin{equation}
\eta=\varepsilon-\varepsilon^0(P).
\label{eq:SM_eta_def}
\end{equation}
Near the stress-free branch, the same zero-field free-energy surface can be written as
\begin{equation}
f_{\rm eff}(P,\varepsilon)
=
f_{\sigma=0} (P)
+\frac{1}{2}C^{P}(P)\eta^2
+\frac{1}{6}D^{P}(P)\eta^3+\cdots,
\label{eq:SM_branch_expansion}
\end{equation}
where $f_{\sigma=0} (P)=f_{\rm eff}[P,\varepsilon^0(P)]$.
The local harmonic form is justified by the near-linear stress-elastic-strain response over the small strain range shown in Fig.~\ref{fig:figs5}.
\begin{equation}
C^{P}(P)
=
\left.
\frac{\partial^2 f_{\rm eff}}{\partial\varepsilon^2}
\right|_{P,\varepsilon^0(P)}.
\label{eq:SM_local_CP}
\end{equation}
The coefficient $D^P(P)$ is the corresponding third derivative with respect to $\varepsilon$ on the same branch.
It describes nonlinear elasticity away from the small elastic-strain region and is not needed for the main-text extraction of $b(P)$.
All quantities in this section refer to the one-dimensional reduced surface $f_{\rm eff}(P,\varepsilon)$ introduced above.

\subsection{What the constant-$Q$ approximation assumes}
Differentiating Eq.~\eqref{eq:SM_sigma0_branch} gives, for the one-dimensional reduced problem,
\begin{equation}
b(P)
\equiv
\frac{\mathrm{d}\varepsilon^0}{\mathrm{d}P}
=
-
\left.
\frac{f_{P\varepsilon}}{f_{\varepsilon\varepsilon}}
\right|_{\varepsilon=\varepsilon^0(P)}
=
\frac{a(P)}{C^P(P)},
\label{eq:SM_b_branch}
\end{equation}
with $a(P)=-f_{P\varepsilon}$ evaluated on the stress-free branch.
The curvature-defined electrostrictive coefficient is
\begin{equation}
Q(P)=\frac{1}{2}\frac{\mathrm{d}b}{\mathrm{d}P}
=
\frac{1}{2}\frac{\mathrm{d}^2\varepsilon^0}{\mathrm{d}P^2}.
\label{eq:SM_Q_branch}
\end{equation}
The conventional constant-electrostriction estimate corresponds to the local approximation
\begin{equation}
\varepsilon^0(P)\approx Q^{(0)}P^2,
\qquad
b(P)\approx 2Q^{(0)}P,
\label{eq:SM_constantQ_limit}
\end{equation}
where $Q^{(0)}=Q(P=0)$ is the small-polarization curvature near the reference state.
It is accurate when the branch remains nearly quadratic over the polarization range of interest.

To illustrate the limited range of the constant-$Q$ approximation, we fitted the stress-free strain branch $\varepsilon^0_{33}(P_3)$ using polynomial forms with different even orders in $P_3$. 
For AlN, the fitted value of the quadratic coefficient depends strongly on the fitting range and on whether higher-order terms are included, reflecting the non-quadratic strain-polarization relation.

\begin{figure}[htbp]
\centering
\includegraphics[width=9cm]{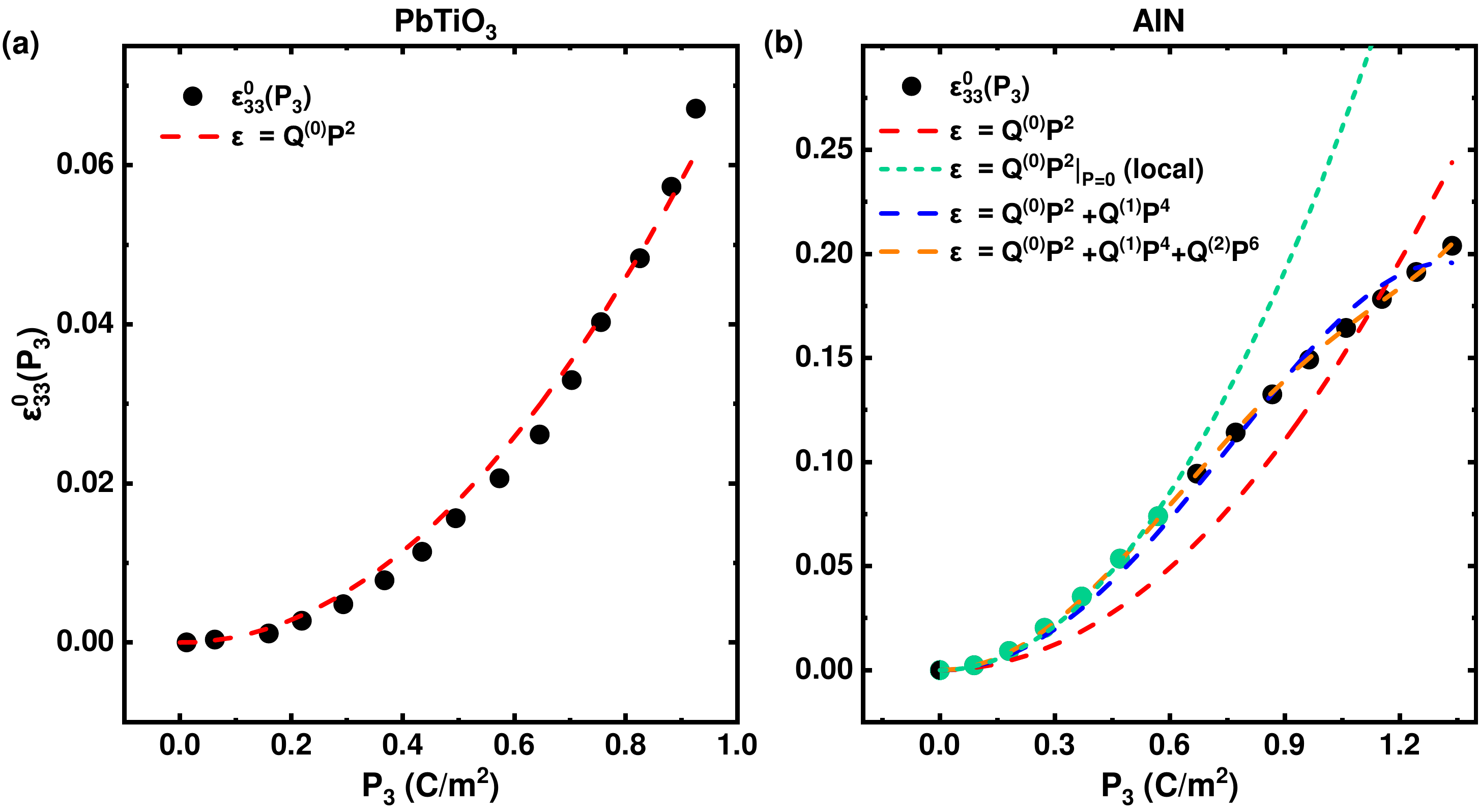}
    \caption{Polynomial representations of the stress-free strain branch $\varepsilon^0_{33}(P_3)$.
    (a) PbTiO$_3$, for which a quadratic fit gives $Q^{(0)}=0.072~\mathrm{m^4/C^2}$. 
    (b) AlN, fitted over different polarization ranges and with polynomial terms of different even orders in $P_3$. 
    The corresponding AlN fitting parameters are listed in Table~\ref{tab:fit_Q}.}
    \label{fig:figS6}
\end{figure}

\begin{table}[htbp]
\centering
\caption{Fitting parameters for different polynomial representations of the stress-free strain branch $\varepsilon^0_{33}(P_3)$ in AlN, corresponding to Fig.~\ref{fig:figS6}(b). Note here subscripts are omitted for simplicity.}
\label{tab:fit_Q}
\begin{tabular}{lccc}
\toprule
\multicolumn{1}{c}{Form} &
\multicolumn{3}{c}{Parameters} \\
\cmidrule(lr){2-4}
&
$Q^{(0)}$ $(\mathrm{m^4/C^2})$ &
$Q^{(1)}$ $(\mathrm{m^8/C^4})$ &
$Q^{(2)}$ $(\mathrm{m^{12}/C^6})$ \\
\midrule
$\varepsilon^0 = Q^{(0)} P^2$
& 0.137 & -- & -- \\
$\varepsilon^0 = Q^{(0)} P^2$ (local, near $P=0$)
& 0.237 & -- & -- \\
$\varepsilon^0 = Q^{(0)} P^2 + Q^{(1)} P^4$
& 0.225 & $-0.0647$ & -- \\
$\varepsilon^0 = Q^{(0)} P^2 + Q^{(1)} P^4 + Q^{(2)} P^6$
& 0.269 & $-0.147$ & 0.0341 \\
\bottomrule
\end{tabular}
\end{table}

\subsection{Connection to Landau language}
For an expansion about the centrosymmetric reference state, scalar quantities such as $C^P(P)$ may be represented by even powers of $P$,
\begin{equation}
C^{P}(P)=C^{(0)}+C^{(1)}P^2+C^{(2)}P^4+\cdots.
\label{eq:SM_poly_C}
\end{equation}
Such a series is a valid way to parameterize the local stiffness in Eq.~\eqref{eq:SM_local_CP}.
In the present work, however, $b(P)$ and $Q(P)$ are obtained directly as local derivatives of $f_{\rm eff}(P,\varepsilon)$ along the stress-free branch.
Thus the breakdown of the constant-$Q$ estimate is best viewed as the limited range of a reference-state constant-coefficient approximation.
If the same surface is represented by a Landau-type series, the observed variation of $C^P(P)$, $a(P)$, and $Q(P)$ would be absorbed into additional symmetry-allowed coefficients.

%